\documentclass[prb, twocolumn,superscriptaddress]{revtex4-2}
\newcommand{\abs}[1]{|#1|}

\DeclareFontFamily{OMS}{oasy}{\skewchar\font48 }
\DeclareFontShape{OMS}{oasy}{m}{n}{%
         <-5.5> oasy5     <5.5-6.5> oasy6
      <6.5-7.5> oasy7     <7.5-8.5> oasy8
      <8.5-9.5> oasy9     <9.5->  oasy10
      }{}
\DeclareFontShape{OMS}{oasy}{b}{n}{%
       <-6> oabsy5
      <6-8> oabsy7
      <8->  oabsy10
      }{}
\DeclareSymbolFont{oasy}{OMS}{oasy}{m}{n}
\SetSymbolFont{oasy}{bold}{OMS}{oasy}{b}{n}

\DeclareMathSymbol{\smallleftarrow}     {\mathrel}{oasy}{"20}
\DeclareMathSymbol{\smallrightarrow}    {\mathrel}{oasy}{"21}
\DeclareMathSymbol{\smallleftrightarrow}{\mathrel}{oasy}{"24}

\usepackage{amsmath, amssymb, amsthm, mathtools, relsize, multirow, enumitem, dsfont}
\usepackage{graphicx,xcolor}
\usepackage{appendix}
\usepackage{ulem}
\usepackage{hyperref}
\usepackage{float}
\usepackage{dsfont}

\definecolor{byzantine}{rgb}{0.56, 0.0, 1.0}

\begin{document}
\title{Periodic strings: a mechanical analogy to photonic and phononic crystals}

\author{R. S. Pitombo}
\email{rs.pitombo@unesp.br}
\affiliation{Instituto de Física Teórica, UNESP - Univ. Estadual Paulista, ICTP South American Institute for
Fundamental Research, Rua Dr. Bento Teobaldo Ferraz 271, 01140-070, São Paulo, SP, Brazil}

\author{M. Vasconcellos}
\email{mateusvsiqueira@gmail.com}
\affiliation{Instituto de Física, Universidade Federal Fluminense, 24210-346, RJ, Brazil}

\author{P. P. Abrantes}
\email{ppabrantes91@gmail.com}
\affiliation{Departamento de Física, Universidade Federal de São Carlos, Rodovia Washington Luís, km 235 - SP-310, 13565-905, São Carlos, SP, Brazil}

\author{Reinaldo de Melo e Souza}
\email{reinaldos@id.uff.br}
\affiliation{Instituto de Física, Universidade Federal Fluminense, 24210-346, RJ, Brazil}

\author{G. M. Penello}
\email{gpenello@if.ufrj.br}
\affiliation{Instituto de Física, Universidade Federal do Rio de Janeiro, 21941-972, RJ, Brazil}

\author{C. Farina}
\email{farina@if.ufrj.br}
\affiliation{Instituto de Física, Universidade Federal do Rio de Janeiro, 21941-972, RJ, Brazil}


\begin{abstract}

We study a periodic vibrating string composed of a finite sequence of string segments connected periodically, with each segment characterized by a constant linear mass density. The main purpose is to provide a configuration that can mimic the properties of photonic or phononic crystals and could be implemented in undergraduate physics laboratories. We demonstrate that this configuration displays frequency intervals for which wave propagation is not allowed (frequency bandgaps), in close analogy to photonic and phononic crystals. We discuss the behavior of these bandgaps when varying physical parameters, such as the values of the linear mass densities, the oscillation frequency, and the number of strings constituting the entire system. Some analogies with the propagation of electronic waves through a crystal lattice in condensed matter physics are also explored.

\end{abstract}

\maketitle

\section{Introduction}

First proposed in 1987 by Yablonovitch \cite{Yablonovitch1987} and John \cite{John1987}, a photonic crystal is an arrangement of different material media so that its effective optical properties are periodic. This configuration can be achieved by a sequence of periodic structures of equally spaced materials with alternating dielectric constants. Historically, however, the study of one-dimensional periodic structures dates back to 1887 with Lord Rayleigh \cite{Rayleigh1888}, who analyzed an experiment performed by Stokes and showed the existence of high reflectivity of light over a narrow well-defined frequency range.

When the period of the spatial modulation is comparable to the wavelength of the light propagating through the material, photonic bandgaps appear -- frequency ranges in which light propagation is forbidden in certain directions \cite{JoannopoulosBook, Schulkin2004, Szmulowicz2004, Guo2006, vonFreymann2013, Cerjan2017}. These are the photonic crystals, electromagnetic analogs to crystalline solids in which forbidden energy bandgaps appear \cite{AshcroftBook}. 
An essential aspect is that these bandgaps heavily depend on the spatial pattern, leading to several strategies for manufacturing photonic crystals \cite{Ryu2017, Wang2020-1}. For further study concerning other properties and numerical modeling of these materials, as well as experimental details on their fabrication and characterization, see Refs. \cite{JoannopoulosBook, vonFreymann2013} and references therein.

Since their discovery, photonic crystals have offered a remarkable chance for light manipulation, attracting much research interest. Because of their capacity to confine and control light of an arbitrary wavelength, such materials have allowed for a variety of technological applications beyond the field of optical physics, such as in optoelectronics \cite{Joannopoulos1997}, displays \cite{Arsenault2007}, sensors \cite{Xu2013}, solar cells \cite{Bermel2007}, light-emitting diodes \cite{Matioli2010}, optical fibers \cite{CerqueiraJr2010}, and in the construction of high-efficiency reflectors \cite{Bruyant2003}. In nature, photonic crystals are also present from inorganic opals \cite{Aguirre2010} to different organic structures in butterfly wing scales, beetle scales, and bird feathers \cite{Zhang2013}.

Similarly, the propagation of elastic or acoustic waves in periodic composite media also produces bandgaps, which gave rise to the concept of phononic crystals, initially proposed in 1992 by Sigalas and Economou \cite{Sigalas1992} and later in 1993 by Kushwaha and collaborators \cite{Kushwaha1993}. Since then, such systems have been extensively studied, presenting different applications in engineering and applied physics, such as in vibration reduction \cite{Richards2003} and noise control \cite{Sanchez-Dehesa2011}. Recent reviews covering these periodic structures can be found in \cite{Hussein2014, Wang2020-2, Liu2020}. Similar phenomenology can be found in semiconductors heterostructures, as can be appreciated in Ref. \cite{Germano2019} and references therein.

Therefore, the investigation of optical and acoustic responses of periodic structures plays a key role in current research topics. This study involves techniques and concepts rarely discussed in standard undergraduate textbooks. An alternative that has been widely used to emulate one-dimensional photonic crystals, both theoretically and experimentally, involves coaxial cables \cite{Hache2002, Sanchez-Lopez2003, Hache2004, Boudouti2007-1, Boudouti2007-2, Perrier2021}. However, the essential aspect for the appearance of bandgaps is the interference of waves, thereby appearing in any wave phenomena. In this regard, we propose the study of a conceptually simpler configuration: the periodic string, which requires only an introductory classical mechanics background. Such a configuration consists of a finite sequence of string segments connected, presenting periodically alternating linear mass densities, which play the role of dielectric permittivities (densities) in photonic (phononic) crystals. This analogy is possible because the light (sound) velocity in a given material is a function of its permittivity (density), while the wave velocity in a string depends on its linear mass density. As in the context of photonic crystals, the bandgaps appear because periodicity implies a destructive interference for waves with frequencies in ranges denominated frequency bandgaps. Pedagogically, it is enlightening to see these bandgaps originating from the interference of waves in a string instead of the more abstract electromagnetic waves in material media. Formally, periodicity would require an infinite set of strings, but we show that, for realistic values of linear mass densities for the strings, only a few are enough to observe the bandgaps. In this way, this setup offers an experimental appeal since the properties of waves in a vibrating string can be observed in an undergraduate physics laboratory. Throughout this paper, we will refer more to photonic crystals, but the analogies can be applied to phononic crystals as well. 

\section{Methodology}
\label{SecMethodolody}

In this section, we introduce the methodology of the transfer matrix to study the periodic strings.
In addition to being easy to implement numerically, this method is powerful to study one-dimensional wave propagation and is a convenient tool to deal with stratified media. It allows for the exact computation of transmittance and reflectance in non-trivial configurations \cite{Walker1994, Balili2012, Zhan2013, Haddouche2017}, including in the description of photonic crystals \cite{Zi1998, Sanchez-Lopez2003}. For an introduction to the transfer matrix formalism in the contexts of photonics, plasmonics, and condensed matter systems, see Refs. \cite{SoukoulisBook, NunoBook, PereyraBook}.

\begin{figure}[b!]
    \centering
    \includegraphics[width=7.9cm]{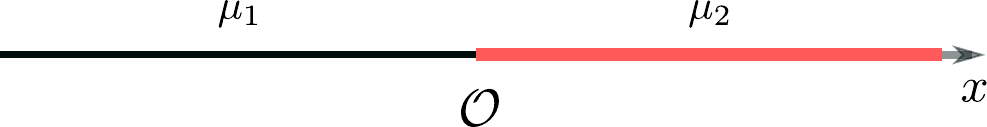}
    \caption{Two semi-infinite strings with different linear mass densities $\mu_1$ and $\mu_2$.}
    \label{fig:semi}
\end{figure}

The starting point is to understand how waves behave at the interface of two different media. Consider the system depicted in Fig. \ref{fig:semi}, consisting of two semi-infinite strings attached at the origin, whose linear mass densities are $\mu_1$ and $\mu_2$. Denoting the vertical displacement of a generic point of the string at instant $t$ by $u(x,t)$, and assuming small slopes ($|\partial u/\partial x| \ll 1$), $u (x, t)$ obeys the wave equation
\begin{equation}
    \frac{\partial^2u(x,t)}{\partial x^2} - \frac{\mu(x)}{T}\frac{\partial^2 u(x,t)}{\partial t^2} = 0 \,,
\label{waveeq}
\end{equation}

\noindent where the tension $T$ will be considered as uniform throughout the string, and $\mu(x)$ is the linear mass density. We first assume that $\mu (x) = \mu_1$ for $x < 0$, and $\mu(x) = \mu_2$ for $x > 0$. Physical solutions must satisfy the boundary conditions \cite{FrenchBook}
\begin{align}
    u_1(0^-,t) &= u_2(0^+,t) \,,
    \label{cc1}
    \\
    \frac{\partial u_1}{\partial x}(0^-,t) &= \frac{\partial u_2}{\partial x}(0^+,t) \,,
    \label{cc2}
\end{align}
where the index $1$ ($2$) refers to $x<0$ ($x>0$).

In this work, we are mainly interested in transmittance and reflectance. To compute them, we search for harmonic solutions of frequency $\omega$ for Eq. (\ref{waveeq}) submitted to the boundary conditions (\ref{cc1}) and (\ref{cc2}). We propose the {\it ansatz}
%
\begin{align}
    & u_1(x,t)=A_1e^{i(k_1x-\omega t)}+B_1e^{-i(k_1x+\omega t)} \;\;\;\;(x<0) \,, \\
    & u_2(x,t)=A_2e^{i(k_2x-\omega t)}+B_2e^{-i(k_2x+\omega t)} \;\;\;\;(x>0) \,,
\end{align}
%
where $A_i$ and $B_i$ ($i=1,2$) are coefficients to be determined. The wave equation requires that
\begin{equation}
    k_j = \omega\sqrt{\frac{\mu_j}{T}} \,, \;\;\;\; \textrm{with } j=1,2 \,,
    \label{disp}
\end{equation}

\noindent and the boundary conditions give the relations
\begin{align}
    & A_1+B_1=A_2+B_2 \,,\\
    & k_1(A_1-B_1)=k_2(A_2-B_2) \,.
\end{align}

\noindent Defining $\eta = k_2/k_1$, this linear system of equations can be cast into the matrix form
\begin{align}
        \begin{pmatrix}
        A_1\\
        B_1
        \end{pmatrix}
        &=
        \frac{1}{2}
        \begin{pmatrix}
        1+\eta & 1-\eta\\
        1-\eta & 1+\eta
        \end{pmatrix}
        \begin{pmatrix}
        A_2\\
        B_2
        \end{pmatrix} =
        \mathds{T}_{1\rightarrow2}
        \begin{pmatrix}
        A_2\\
        B_2
        \end{pmatrix}. 
        \label{transf}
    \end{align}
\noindent $\mathds{T}_{1 \rightarrow 2}$ is called the transfer matrix \cite{NunoBook}, and it relates the coefficients of the solutions at the interface that connects the waves traveling in the left and right sides of the origin. Thus, we can obtain the reflectance and transmittance directly from the transfer matrix elements, enforcing that no waves are coming from the right and setting $B_2=0$. Since the transmittance (reflectance) is the ratio between the power carried by the transmitted (reflected) and incident waves, we can write
%
\begin{align}
         &\mathcal{T} = \frac{k_2}{k_1} \frac{\abs{A_2}^2}{\abs{A_1}^2} \;\;\; \textrm{and} \;\;\; \mathcal{R} = 
         \frac{\abs{B_1}^2}{\abs{A_1}^2} \, .
\end{align}

\noindent From Eq. (\ref{transf}), we obtain $2A_1=(1+\eta)A_2$, implying
%
%
%
%
\begin{align}
    \mathcal{T} =\frac{4k_2/k_1}{\left(1+k_2/k_1\right)^2} = \frac{4\sqrt{\mu_2/\mu_1}}{\left(1+\sqrt{\mu_2/\mu_1}\right)^2} \,.
\end{align}
    
\noindent Analogously, the reflectance reads
\begin{equation}
\mathcal{R} = \bigg(\frac{1-k_2/k_1}{1+k_2/k_1}\bigg)^2 =\bigg(\frac{1-\sqrt{\mu_2/\mu_1}}{1+\sqrt{\mu_2/\mu_1}}\bigg)^2 \,. \label{R}    
\end{equation}

\begin{figure}[t!]
    \centering
    \includegraphics[width =7.9cm]{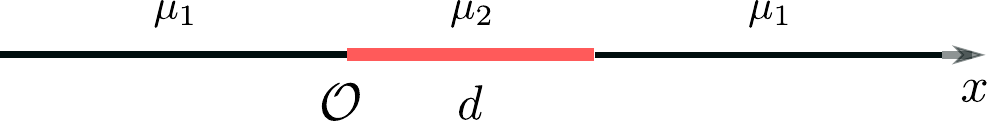}
    \caption{Finite string segment with length $d$ and linear mass density $\mu_2$ placed between two semi-infinite strings of equal linear mass density $\mu_1$.}
    \label{fig:segFin}
\end{figure}

Let us now analyze the case where a finite string segment of length $d$ and linear mass density $\mu_2$ is inserted between two semi-infinite strings of equal linear mass densities $\mu_1$, as in Fig. \ref{fig:segFin}. The transmittance and reflectance can be obtained through the total transfer matrix, relating the amplitudes of the solutions in the intervals $x<0$ and $x>d$. The additional phase acquired by the propagation of the waves along the finite string segment can be taken into account by the propagation matrix
\begin{equation}
    \mathds{P}_2=
        \begin{pmatrix}
        e^{ik_2d} & 0\\
        0 & e^{-ik_2d}
        \end{pmatrix}
        \label{Prop}
\end{equation}

\noindent so that the total transfer matrix is
\begin{align}
    \mathds{T}_{\mathrm{tot}}&=\mathds{T}_{1\rightarrow2}\mathds{P}_2\mathds{T}_{2\rightarrow1} \,,
\label{ttot}
\end{align}

\noindent with $\mathds{T}_{2\rightarrow 1} = \mathds{T}^{-1}_{1\rightarrow 2}$ by definition. From the total transfer matrix, the transmittance and reflectance are obtained in the same way as in the previous example, yielding
\begin{align}
   \mathcal{T}&=\frac{k_f}{k_i}\frac{1}{|T_{11}^{\mathrm{tot}}|^2}\,,
   \label{Transm} \\
   \mathcal{R}&= \bigg|\frac{T_{21}^{\mathrm{tot}}}{T_{11}^{\mathrm{tot}}}\bigg|^2 \,,
    \label{Refl}
\end{align}

\noindent where $k_i$ ($k_f$) stands for the wavenumber in the semi-infinite segment in the range $x<0$ ($x>d$) and $T_{ij}^{\rm tot}$ ($i, j = 1, 2$) are the matrix elements of $\mathds{T}_{\mathrm{tot}}$. In the particular configuration of Fig. \ref{fig:segFin}, the linear mass densities of these segments were chosen to be the same, and, consequently, $k_f/k_i = 1$. Explicit calculation (see Appendix \ref{AppendA}) results in
	 \begin{equation}
	     \mathcal{T} = \frac{1}{1+\dfrac{(\mu_1^2-\mu_2^2)^2}{\mu_1^2\mu_2^2}\sin^2\left({\omega d\sqrt{\dfrac{\mu_2}{T}}}\right)} \,.
	     \label{transm1seg}
	 \end{equation}

	 \begin{figure}[t!]
	     \centering
	     \includegraphics[width=8.1cm]{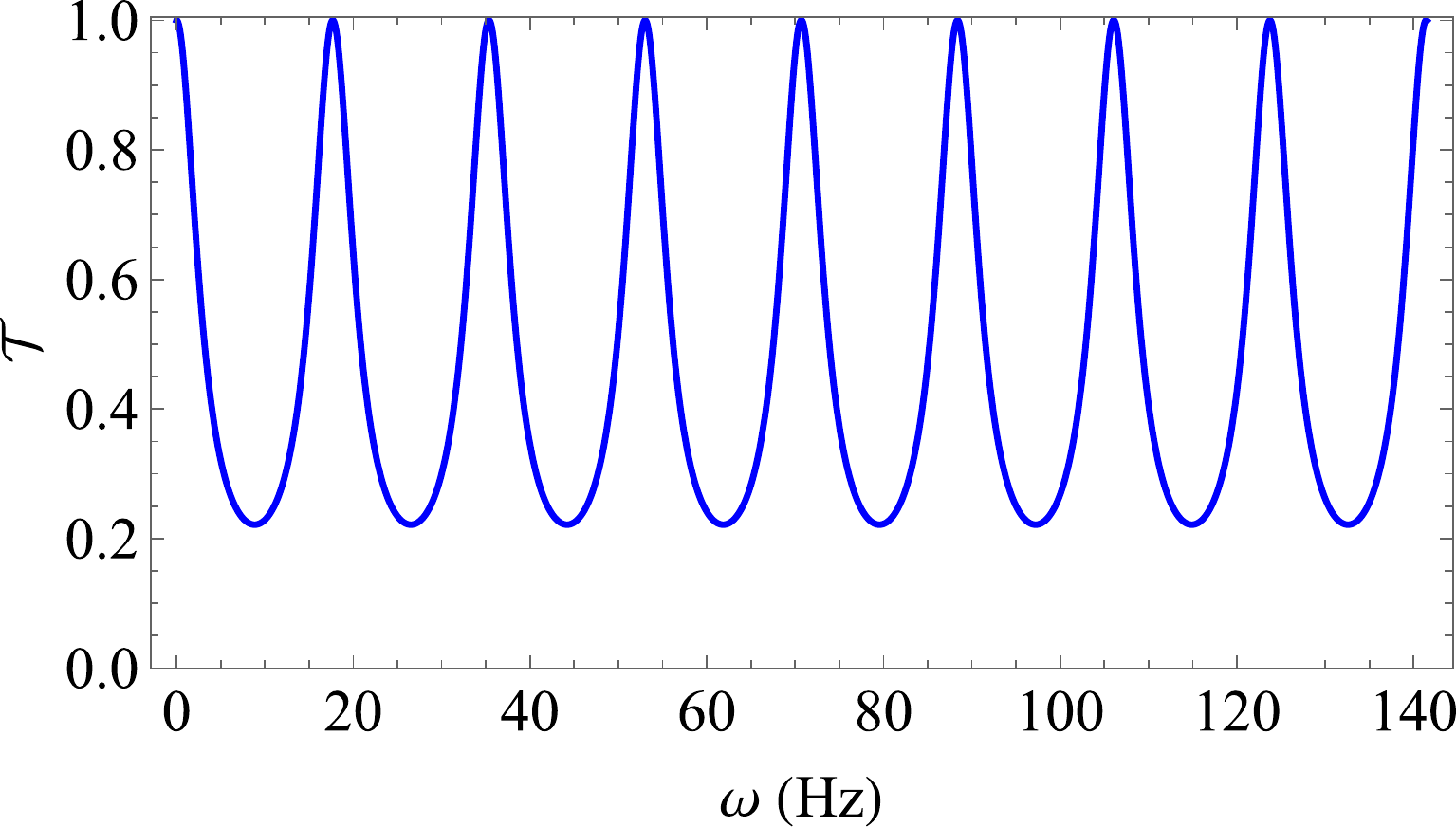}
	     \caption{Transmittance $\mathcal{T}$ as a function of frequency $\omega$ for scattering by a single finite segment. Here, $\mu_1=2$~g/m, $\mu_2 = 4\mu_1$, $d = 10$ cm, and $T=1$ N.}
	     \label{fig:n1}
	 \end{figure}

This expression is plotted in Fig. \ref{fig:n1}, in which we show the transmittance as a function of frequency, with $\mu_1=2$~g/m, $\mu_2 = 4\mu_1$, $d = 10$ cm, and $T=1$ N. Note that the transmittance never vanishes and presents a periodic behavior with a period satisfying $\Delta\omega d\sqrt{\mu_2/T}=\pi$. From Eq. (\ref{disp}), we 
obtain $\Delta k_2 d = \pi$. Equations (\ref{Prop}) and (\ref{ttot}) show that, when this condition is satisfied, $T_{11}^{\rm tot}$ just changes sign, keeping $\mathcal{T}$ unaltered. Also, whenever $k_2 d = n\pi$ with integer $n$, we have full transmittance. This was also expected from Eqs. (\ref{Prop}) and (\ref{ttot}) since, in this case, $\mathds{P}_2$ and, consequently, $\mathds{T}_{\rm tot}$ are $\pm 1$ times the identity matrix. We invite the reader to explain these features from the interference of multiple reflections and check from themselves how time-saving the transfer matrix method is. Note that, in the limit $\mu_1\rightarrow\infty$ for a fixed $\mu_2$, we have $\mathcal{T}\rightarrow 0$, except at the frequencies for which we have full transmittance. There is a close analogy between this case and the transmission of electromagnetic waves propagating normally through a finite dielectric layer. The functional dependence is the same, with the only differences being the relevant physical parameters. As a matter of fact, the role played by linear mass densities in our cases is analogous to the one played by the refraction indexes in the electromagnetic one.
	 
\begin{figure}[h!]
	     \centering
	     \includegraphics[width=7.9cm]{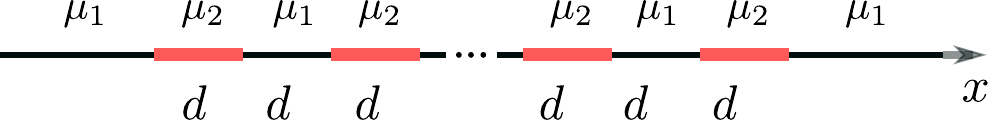}
	     \caption{The periodic string: a set of $N$ segments with linear mass density $\mu_2$ and size $d$ alternated with segments of linear mass density $\mu_1$ and the same size.}
	     \label{fig:cperiod}
	 \end{figure}

Now, we return to the central purpose of this work: a periodic string containing $N$ segments of linear mass density $\mu_2$ interspersed with segments characterized by linear mass density $\mu_1$, as sketched in Fig. \ref{fig:cperiod}. By a straightforward generalization of the previous examples, the transfer matrix is now given by
\begin{equation}
    \mathds{T}_{\mathrm{tot}}=\mathds{T}_{1\rightarrow2} \left( \mathds{P}_2\mathds{T}_{1\rightarrow2}^{-1}\mathds{P}_1\mathds{T}_{1\rightarrow2} \right)^{N-1}\mathds{P}_2\mathds{T}_{1\rightarrow2}^{-1} \, ,
\label{EqPS}
\end{equation}
where $\mathds{P}_1$ accounts for the propagation along segments of linear mass density $\mu_1$ and is obtained from Eq. (\ref{Prop}) by exchanging $k_2$ by $k_1$ (we shall assume all segments with length $d$). In the next section, we study Eq. (\ref{EqPS}) analytically, but it is instructive for the students to gain some intuition by first analyzing some concrete examples. For the results of Sec. \ref{SecResults}, we used the computational environment Mathematica, but no specific calculation techniques or any numerical approximations were needed. Additionally, an experimental proposal is discussed in Appendix \ref{AppendB}.
	  
\section{Results and discussions}
\label{SecResults}

\begin{figure}[t!]
    \centering
    \includegraphics[width=7.9cm]{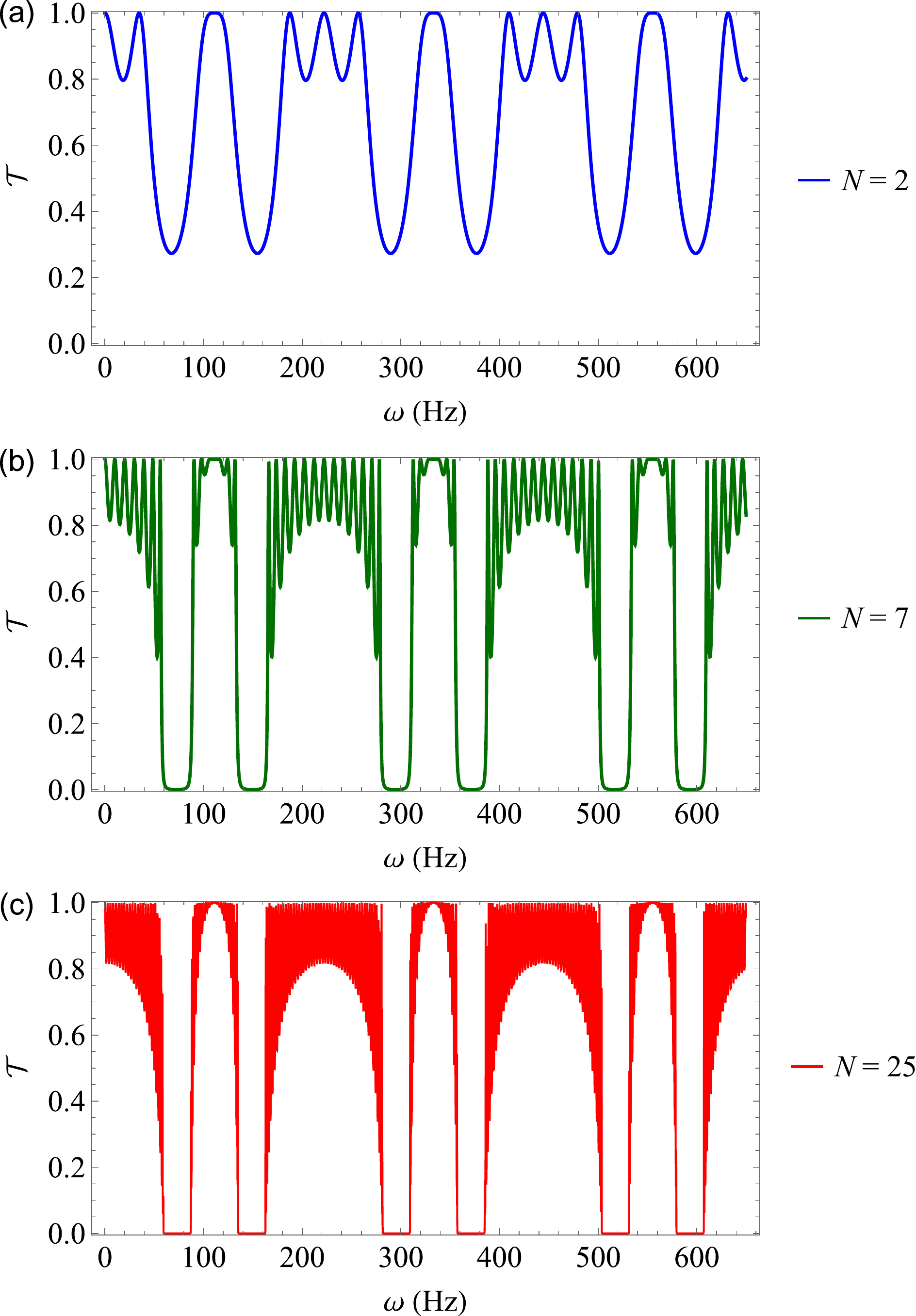}
    \caption{Transmittance $\mathcal{T}$ as a function of the frequency $\omega$ for {\bf (a)} $N = 2$, {\bf (b)} $N = 7$, and {\bf (c)} $N = 25$. In all plots, we set $\mu_1=2$~g/m and $\mu_2 = 4\mu_1$.
    }
    \label{freq}
\end{figure}

Throughout this section, we chose $d=10$~cm and $T=1$~N. In Fig. \ref{freq}, we show the transmittance against the frequency of the propagating wave along the string. In all panels, we used $\mu_1=2$~g/m and $\mu_2=4\mu_1$, and each curve refers to a configuration with a different value of $N$, the number of unit cells contained in the string. A striking feature in Figs. \ref{freq}(b) and \ref{freq}(c) is the presence of periodic frequency regions in which the transmittance goes to zero, revealing the presence of bandgaps. Note that we still do not have bandgaps for $N=2$ [panel (a)], although the transmittance is already much lower for some frequencies. As $N$ increases, the valleys become more prominent, and bandgaps can be identified even for a modest value of $N=7$ [panel (b)]. In addition, between these bandgaps, there is an oscillatory behavior, and the frequency of such oscillations increases with $N$.

\begin{figure}[t!]
    \centering
    \includegraphics[width=8.1cm]{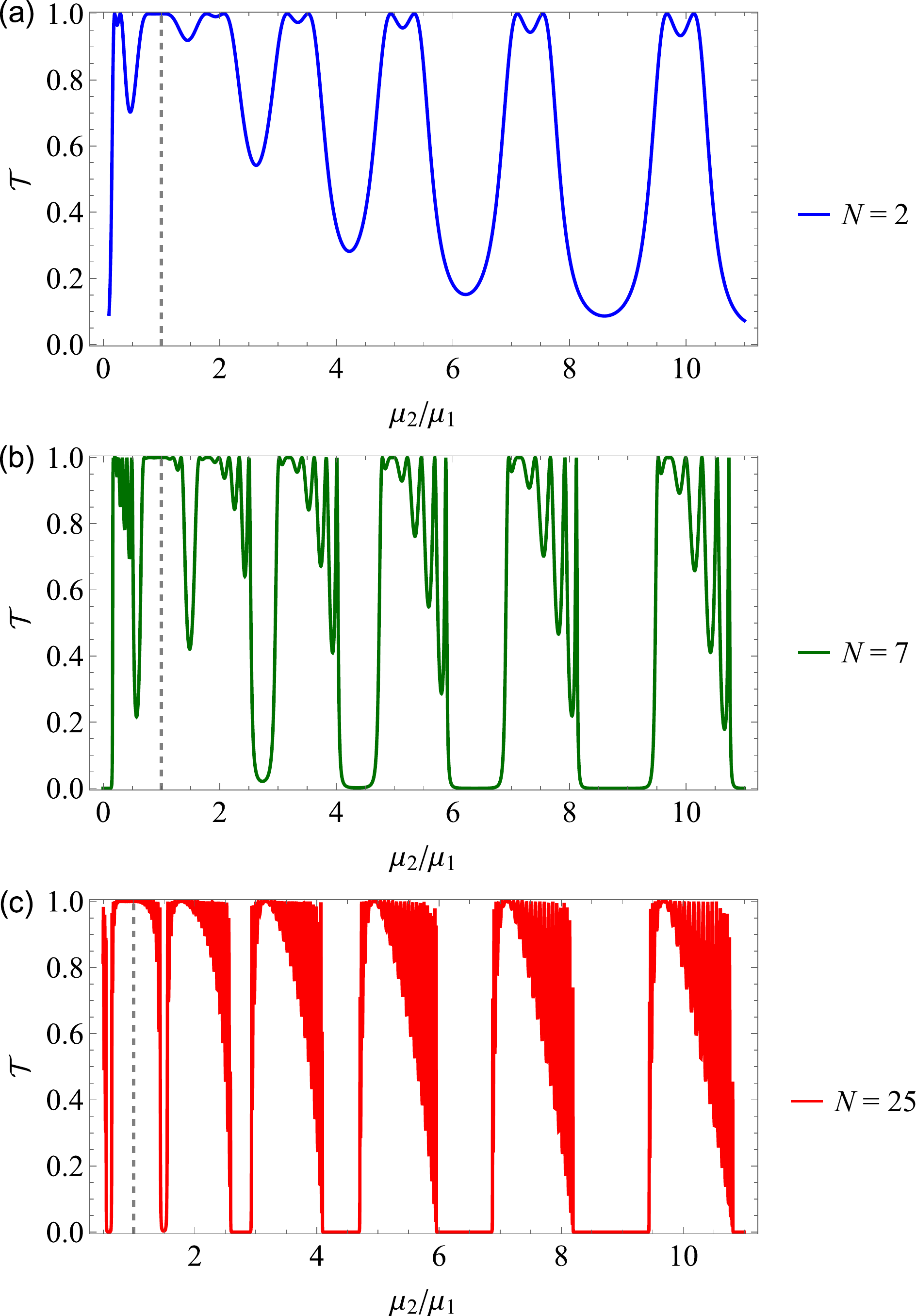}
    \caption{Transmittance $\mathcal{T}$ as a function of the strings' density ratio $\mu_2/\mu_1$ for {\bf (a)} $N = 2$, {\bf (b)} $N = 7$, and {\bf (c)} $N = 25$. The oscillation frequency was fixed at $\omega = 500$ Hz and we used $\mu_1=2$~g/m.}
    \label{dens}
\end{figure}

Figure \ref{dens} displays the transmittance against the density ratio $\mu_2/\mu_1$ of the strings segments, with $\mu_1=2$~g/m and exploiting the same values of $N$ shown in Fig. \ref{freq}. The dashed gray lines in each panel stress that the outcome is full transmittance for $\mu_2/\mu_1 = 1$, as expected for a homogeneous medium. These plots indicate that, for a given frequency, we can always choose a density ratio value for which that frequency is contained in a bandgap.

\begin{figure}[t!]
    \centering
    \includegraphics[width=8.1cm]{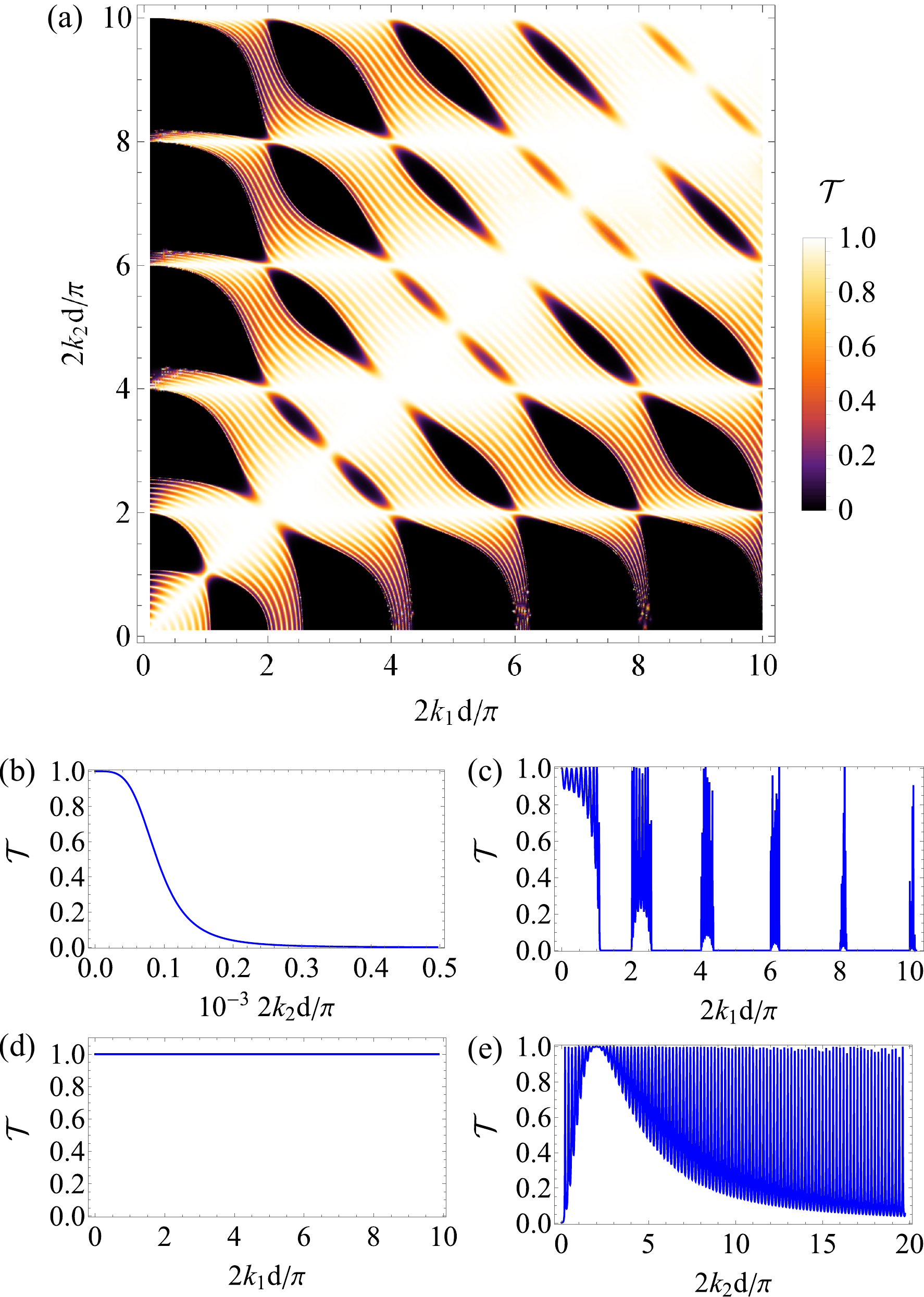}
    \caption{ Transmittance $\mathcal{T}$ as a function of {\bf (a)}  $k_1$ and $k_2$, {\bf (b)} $k_2$, for the slope $k_2\gg k_1$, {\bf (c)} $k_1$, for the slope $k_2\ll k_1 $, {\bf (d)} $k_1$, with $k_2=\pi/d$ and fixed frequency, and {\bf (e)} $k_2$, with $k_1=\pi/d$ and fixed frequency. In all plots, we set $N = 10$ and the axes with $k_1$ and $k_2$ are normalized by $\pi/2d$ for convenience.}
    \label{k1k2}
\end{figure}

In Fig. \ref{k1k2}(a), we display the transmittance as a function of the wavenumbers of each string for $N = 10$. Consider straight lines passing through the origin, with a constant slope given by the ratio $k_2/k_1 = \sqrt{\mu_2/\mu_1}$, according to Eq. (\ref{disp}). For convenience, but without loss of generality, suppose that both $\mu_1$ and $\mu_2$ are fixed. In this case, the wavenumber of each string can only change through the frequency of the harmonic wave propagating along the string. Thus, starting at the origin and moving along any of these lines, we continuously increase the frequency of the wave. As it is evident from Fig. \ref{k1k2}(a), the transmittance is always equal to $1$ for $k_2/k_1= 1$, as expected, since it implies $\mu_2 = \mu_1$. However, even for slight deviations of $k_2/k_1 = 1$, we necessarily pass through bandgaps when moving from the origin along a straight line. Note that the crossed bandgaps are very narrow for small deviations of $k_2/k_1 = 1$, but wider bandgaps appear as we deviate from unity. We can understand this result if we remember that there is no reflection of the incident wave for a homogeneous string. Consequently, there is nothing to scatter this wave: nor any discontinuity in the linear mass density of the string and not even any continuous change in density. Nevertheless, as the ratio $\mu_2/\mu_1$ deviates from $1$, the existence of the \lq\lq lattice{\rq\rq} (periodic string) starts to be noticed, and the bandgaps start to show up. In this sense, this ratio measures the presence of the lattice and, consequently, how much scattering occurs in the system. It is worth noting that, even for slopes very different from the unity, besides wide bandgaps, we still have narrow ones, as shown in Fig. \ref{freq}, which can be understood as similar cuts in Fig. \ref{k1k2} with constant slopes containing the origin.

Naively, one could think that Fig. \ref{k1k2}(a) is symmetric concerning the straight line containing the origin with slope $k_2/k_1 = 1$. However, this is not true since our physical system is not symmetric under the exchange $k_2 \leftrightarrow k_1$ (recall that the incident wave starts propagating through a semi-infinite string with linear mass density $\mu_1$ and, after crossing the \lq\lq periodic string{\rq\rq}, is transmitted to another semi-infinite string with the same density $\mu_1$). This asymmetry becomes less evident as the number $N$ of unit cells composing the periodic string increases. 

There are two interesting limiting cases. The first one is $k_2/k_1 \rightarrow \infty$, which can be thought as taking $\mu_2 \rightarrow \infty$ and a finite value for $\mu_1$. This situation is similar to an incident wave reaching a piece of string with infinite mass, which means an incident wave that reaches a fixed extreme (Dirichlet boundary condition). Consequently, we expect total reflection or, equivalently, zero transmittance. Indeed, Fig. \ref{k1k2}(b) shows the transmittance as a function of $k_2$ for the vertical axis containing the origin ($k_2\gg k_1$), and the outcome is zero transmittance. One could think that at $2k_2d/\pi$ equal to an even positive integer, implying $d = n\lambda_2/2$ ($n = 1, 2, 3,...$), the transmittance would not vanish. Nevertheless, even for these values, the transmittance tends to zero as the slope tends to infinite, though more slowly, as the condition $d = n\lambda_2/2$ favors constructive interference [the same kind of constructive interference that we discussed with Eq. (\ref{transm1seg})]. The second limiting case consists of $k_2\ll k_1$, and it corresponds to select in Fig. \ref{k1k2}(a) the horizontal straight line containing the origin, resulting in Fig. \ref{k1k2}(c).

A last interesting analysis comes from considering horizontal and vertical cuts in Fig. \ref{k1k2}(a) that do not include the origin, which means considering constant values of $k_2$ and $k_1$, respectively. Since $k_2 = \omega \sqrt{\mu_2/T}$, a constant $k_2$ means a constant value of $\omega\sqrt{\mu_2}$ (the string tension $T$ is held constant along all this work). As we seek to quantify the influence of a change in the parameters on a given propagating mode, we may consider a fixed frequency, such that constant $k_2$ implies constant $\mu_2$. Consequently, starting at a point in a vertical axis of Fig. \ref{k1k2}(a) and moving along a horizontal straight line is equivalent to increasing the value of $\mu_1$. The first important feature to be noted is that there are no bandgaps for the horizontal lines given by $2k_2d/\pi$ equal to an even positive integer (or $d = n\lambda_2/2$, with $n = 1, 2, 3,...$), as highlighted in Fig. \ref{k1k2}(d). In other words, no bandgap will take place, regardless of the choice of propagation mode frequency and also the value of $\mu_1$. Moreover, $\mathcal{T} = 1$ for these values, which can be understood in terms of the constructive interference condition already mentioned. Analogous reasoning can be employed to the vertical cuts: moving along a straight vertical line means increasing $\mu_2$. For values of $2k_1d/\pi$ equal to an even positive integer (or $d = n\lambda_1/2$, with $n = 1, 2, 3,...$), there is an oscillatory behavior, so that the transmittance is not always equal to one, but again there are no bandgaps, as shown in Fig. \ref{k1k2}(e).


\section{Analytical origin of bandgaps}

\begin{figure}[t!]
    \centering
   \includegraphics[width=7.5cm]{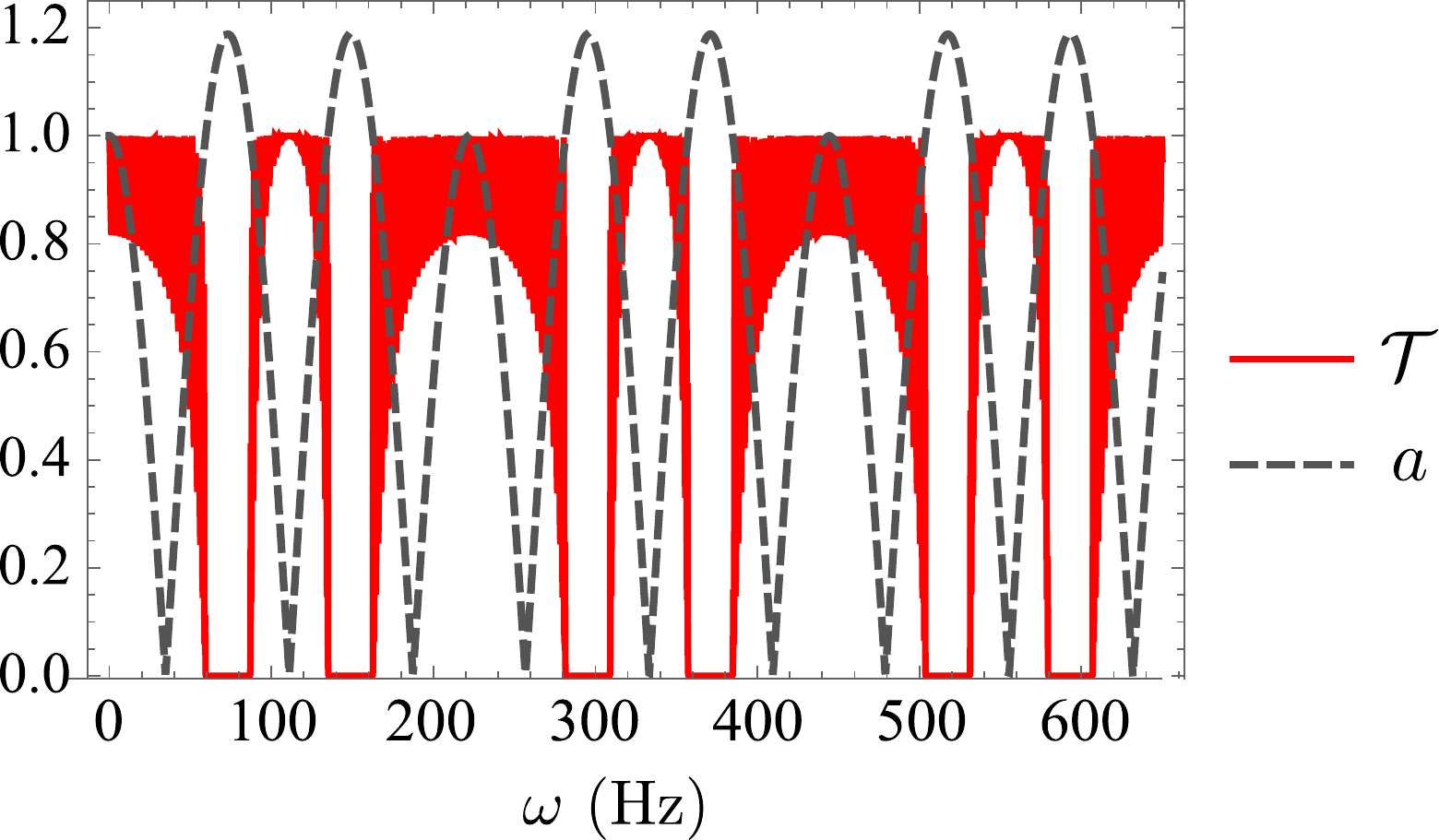}
   \caption{Transmittance $\mathcal{T}$ and the parameter $a$ as a function of the frequency $\omega$ for $\mu_1=2$~g/m, $\mu_2 = 4\mu_1$, and $N = 25$.}
\label{a}
\end{figure}

At this point, one might wonder why the frequency gaps appear and if it is possible to determine their position analytically. The periodicity allows for a positive answer. An analytical evaluation of Eq. (\ref{EqPS}) requires the $(N-1)$-th power of the matrix $\mathds{M}=\mathds{P}_2\mathds{T}_{1\rightarrow2}^{-1}\mathds{P}_1\mathds{T}_{1\rightarrow2}$. Since $\det \mathds{M}=1$, we have \cite{BW}
\begin{equation}
   (\mathds{M})^n =  \begin{pmatrix}
m_{11}U_{n-1}(a)-U_{n-2}(a) & m_{12}U_{n-1}(a) \\ 
m_{21}U_{n-1}(a) & m_{22}U_{n-1}(a)-U_{n-2}(a) \\  
\end{pmatrix}\, , \label{M}
\end{equation}
where $m_{ij}$ denotes the matrix elements of $\mathds{M}$,  $a = (m_{11}+m_{22})/2$, and $U_n(x)$ are the Chebyshev polynomials of the second kind, given by
\begin{equation}
    U_n(x) = \frac{\sin[(n+1)\cos^{-1}(x)]}{\sqrt{1-x^2}} \, .
\end{equation}
Since $\cos^{-1}(x)$ is a purely imaginary number for $|x|>1$, and recalling that $\sin(i\alpha)=-i\sinh(\alpha)$, we conclude that, for $|x|>1$, $U_n(x)$ diverges exponentially for large $n$, leading to a divergence in all elements of $\mathds{T}_{\rm tot}$ and, from Eq. (\ref{Transm}), to a zero transmittance. From Eq. (\ref{M}), we see that $U_n$ must be evaluated at the semi-trace $a$ of the matrix $\mathds{M}$, which is a function of $k_1$, $k_2$ and $d$. Hence, the explicit calculation of $a$ shows that the frequency bandgaps appear whenever the following condition is satisfied:
\begin{equation}
   \frac{\Bigg|2\cos(k_1d)\cos(k_2d)-\left(\dfrac{k_1}{k_2}+\dfrac{k_2}{k_1}\right)\sin(k_1d)\sin(k_2d) \Bigg|}{2} > 1 \, . \label{bandgap}
\end{equation}
Outside this range, $U_n(x)$ is an oscillating function, explaining the oscillations in the transmittance observed in the previous section. In Fig. \ref{a}, we observe that the gaps appear exactly in the regions where $a>1$. 

Although we do not have an explicit solution of Eq. (\ref{bandgap}), its main properties can be readily grasped. First, note that, if $k_1=k_2$ (homogeneous string),  $a$ will become $\cos^2(2kd)$ and the condition (\ref{bandgap}) is never satisfied. Indeed, we have obtained full transmittance for all frequencies. Furthermore, when $k_2$ and $k_1$ are commensurable, $a$ is a periodic function, and the bandgaps' locations and width will obey the same periodicity. When $|k_1-k_2|/k_1 = |\mu_1-\mu_2|/\mu_1 \sim 1$, we have $U_n\sim e^n$ and thus, from Eq. (\ref{Refl}), $\mathcal{R}\sim e^{-n}$. This explains why the bandgaps are perceptible even for small values of $N$ in the previous section.

\section{Final remarks and conclusions}
\label{SecFinalRemarks}

In this work, we employed the transfer matrix formalism to investigate the periodic string and explore its analogies with photonic crystals and condensed matter systems. We determined the transmittance and discussed its dependence on the relevant physical parameters. The most remarkable result was the emergence of prohibited frequency bandgaps, even for a low number of strings segments. The choice of the physical parameters was made to be compatible with materials and equipment that may be available in undergraduate physics laboratories as discussed in appendix B.

Useful analogies may also be drawn from our results to the behavior of band theory in crystalline systems. The interfaces in our string system are analogous to the atoms that scatter the electronic wavefunction, and the frequencies in our classical system play the role of energy, as expected by the de Broglie relation applied to the electronic wavefunction. Also, the expression $|1-\eta| = |1-\sqrt{\mu_2/\mu_1}|$ can be related to the potential generated by each atom. For instance, when $\eta = 1$, we have $\mu_1 = \mu_2$, and there is no scattering, which would be analogous to the free electron model. In fact, as expected, we obtain in Fig. \ref{k1k2} (d) full transmission in this case, tantamount to state that a free electron gas presents no energy gap. In addition, for $|1-\eta|\ll 1$, we have weak scattering, analogously to the nearly free electron approximation, leading to the appearance of small bandgaps. The small reflectances, given in Eq. (\ref{R}), are akin to the small matrix element of the potential connecting states with different wave vectors in the condensed matter case. The opposite situations of $\eta\gg 1$ and $\eta\ll 1$ are analogous to the tight-binding limit. In these limits, we may think of Fig. \ref{fig:segFin} as the analog of a single atom. As we discussed in Sec. \ref{SecMethodolody}, in the limit $\eta\rightarrow 0$ or $1/\eta\rightarrow$ 0, the result is transmission only on a discrete set of frequencies, which plays the role here of the discrete atomic energy levels in crystals. Once we couple many string segments as in Fig. \ref{fig:cperiod} (or \lq\lq atoms\rq\rq in this analogy), the small individual transmittance acts as the hopping parameter, which enlarges the discrete transmitted frequencies into a small band of allowed frequencies. Despite these similarities, the periodic string also has some important differences from Bloch waves in crystals. To begin with, we do not have de Broglie relations so that frequency band does not imply energy bands. In our classical system, any energy is possible. Furthermore, in our classical system, we have an inhomogeneous medium, and thus the wave has different wavelengths depending on which medium it is vibrating. Therefore, our analogy is stronger with the envelope approximation in heterostructures than homogeneous crystals.

Ultimately, this system offers a variety of interesting aspects to explore, providing much physical intuition for being described purely within the framework of classical mechanics. Therefore, it is an excellent route to get beginner undergraduate students in touch with physical concepts rarely covered in standard textbooks. It can also arouse the student's interest in other topics and motivate further studies, as the techniques we used have found applications in different physical situations and can be generalized for more complex problems, as discussed throughout the text.


\section*{Acknowledgements}


R.S.P. and C.F. thank L. O. A. Azevedo for useful discussions. R.S.P. and P.P.A. acknowledges São Paulo Research Foundation (FAPESP) (Grant numbers \#2020/14489-5 and \#2021/04861-7) for financial support. M.V.S acknowledges Coordenação de Aperfeiçoamento de Pessoal de Nível Superior (CAPES) (Grant number 88887.635842/2021-00) for financial support. G.P.M acknowledges Fundação de Amparo à Pesquisa do Estado do Rio de Janeiro (FAPERJ). C.F. acknowledges Conselho Nacional de Desenvolvimento Cient\'{i}fico e Tecnol\'{o}gico (CNPq) (Grant number 310365/2018-0). R.M.S. also thanks the funding agencies.

\vspace{0.1in}



\appendix
\section{Explicit derivation of Eq. (\ref{transm1seg})}
\label{AppendA}

In this appendix, we discuss two strategies to derive the transmittance of the system depicted in Fig. \ref{fig:segFin}, given by Eq. (\ref{transm1seg}). First, we impose the boundary conditions on the harmonic solutions and solve the system of linear equations that arises from these steps. In the second method, we re-obtain this result, employing the transfer matrix formalism and illustrating its convenience.

\subsection{By imposing the boundary conditions at the interfaces}

Supposing that there is an incoming wave in the region $x<0$, the harmonic solutions in each of the segments are 
\begin{align}
    u_1(x,t) &= Ae^{i(k_1x-\omega t)}+B e^{-i(k_1x+\omega t)} \;\; (x<0)\,,
    \\
    u_2(x,t) &= Ce^{i(k_2x-\omega t)}+D e^{-i(k_2x+\omega t)}  \;\, (0<x<d)\,,
    \\
    u_{3}(x,t) &= Ee^{i(k_1x-\omega t)}  \;\;\;\;\;\;\;\;\;\;\;\;\;\;\;\;\;\;\;\;\;\;\,\,\;\;\: (x>d)\,.
\end{align}

\noindent Imposing the boundary conditions at the interfaces $x = 0$ and $x = d$, we obtain set of the equations
\begin{align}
    &A+B=C+D \,, \label{A1}\\
    &k_1(A-B)=k_2(C-D) \,,\label{A2}\\
    &Ee^{ik_1d}=Ce^{ik_2d}+De^{-ik_2d} \,,\label{A3}\\
    &Ek_1e^{ik_1d}=k_2 \left(Ce^{ik_2d}-De^{-ik_2d} \right) \,.\label{A4}
\end{align}

After some simple mathematical manipulations, one can obtain
\begin{align}
    \frac{A}{E} =& \frac{1}{4k_1k_2}\left[(k_1+k_2)^2e^{i(k_1-k_2)d}-(k_1-k_2)^2e^{i(k_1+k_2)d}\right] .
\end{align}

Recalling that
\begin{align}
    \frac{1}{\mathcal{T}}=\frac{|A|^2}{|E|^2}=\frac{AA^*}{EE^*} \,,
\end{align}

\noindent then
\begin{align}
    \frac{1}{\mathcal{T}} &= \frac{1}{16k_1^2k_2^2} \left[(k_1+k_2)^2e^{-ik_2d}-(k_1-k_2)^2e^{ik_2d} \right]\nonumber\\
    &\times\left[ (k_1+k_2)^2e^{ik_2d}-(k_1-k_2)^2e^{-ik_2d} \right] \,. \label{A17}
\end{align}

\noindent After lengthy but straightforward manipulations, we arrive at Eq. (\ref{transm1seg}).


\subsection{By transfer matrix approach}

In this case, it is only necessary to calculate the matrix element $T_{11}^{\mathrm{tot}}$ of the transfer matrix $\mathds{T}_\mathrm{tot}$, given by Eq. (\ref{ttot}). We can write
\begin{align}
    \mathds{T}_\mathrm{tot}&=\frac{1}{2}
        \begin{bmatrix}
        1+\eta & 1-\eta\\
        1-\eta & 1+\eta
        \end{bmatrix}
          \begin{bmatrix}
        e^{ik_2d} & 0\\
        0 & e^{-ik_2d}
        \end{bmatrix}
        \mathds{T}_{1\rightarrow2}^{-1} \label{A21}
\end{align}

\noindent Using that $\mathds{T}_{1\rightarrow2}^{-1}=\mathds{T}_{2\rightarrow1}$, the previous equation yields
\begin{align}
    T_{11}^\mathrm{tot}&=\frac{1}{4} \left[ e^{ik_2d}(1+\eta)(1+\eta^{-1})+e^{-ik_2d}(1-\eta)(1-\eta^{-1}) \right] \cr \cr
    &=\frac{1}{4k_1k_2} \left[ e^{ik_2d}(k_1+k_2)^2-e^{-ik_2d}(k_1-k_2)^2 \right] \,.
\end{align}

\noindent In going to the second line, we used the definition $\eta = k_2/k_1$ given in the main text. Consequently, using Eq. (\ref{Transm}), the transmittance is obtained from
\begin{align}
    \frac{1}{\mathcal{T}} &= \frac{1}{16k_1^2k_2^2} \left[(k_1+k_2)^2e^{-ik_2d}-(k_1-k_2)^2e^{ik_2d} \right]\nonumber\\
    &\times\left[ (k_1+k_2)^2e^{ik_2d}-(k_1-k_2)^2e^{-ik_2d} \right] \,.
\end{align}

\noindent This is precisely the same result given in Eq. (\ref{A17}) that leads to Eq. (\ref{transm1seg}).


\section{Experimental setup proposal}
\label{AppendB}

To show the frequency intervals for which wave propagations are not allowed (the frequency bandgaps), an adaptation of Melde’s original experiment\cite{melde} can be employed. Modern versions of this experiment can be set up with an electric vibrator, string, pulley, and hanging masses to control the string tension. Also, low-cost experiments are readily available in the literature \cite{Bozzo2019}. The basis of our proposed experiment is similar to Melde’s apparatus, with the main difference being that the string will be replaced by two others with different linear mass densities interspersed. Since we want to observe the forbidden propagation for certain frequencies, the lack of oscillations in the last segment (the one away from the oscillator) will demonstrate that no wave has been transmitted through the stratified medium. In other words, unlike Melde's original experiment that looks for stationary waves, we will focus our attention on whether the wave propagation reaches the end of the string. 

To assemble the experiment, we suggest using an electric vibrator, 
a signal generator to drive the vibrator, 
two strings with different linear mass densities (paracord type IA and paracord type II, for example), and hanging masses to control the tension. The values for frequencies and linear mass densities in the main text were chosen based on typical values obtained with available materials.
There are three different alternatives to observe the lack of oscillations on the last segment of the string: the use of a high-speed camera, the use of stroboscopic light, or the use of opto-switch sensors to measure the motion \cite{LeCarrou2014}.

The choice of paracords (or any nylon-based cord) is due to their ability to melt when exposed to heat. By cutting two segments of different paracords, one can join them seamlessly, making the interface abrupt and respecting the boundary conditions used in the theoretical modeling. This joining procedure must be repeated for each segment, obtaining, in the end, the desired stratified medium with a specific number of segments. A hanging object of mass 100g suspended can be used to create a tension of $1$ N. Considering seven repetitions to generate the situation of Fig. \ref{freq}(b), the total length of the interspersed ropes is $1.4$ m. A $1.0$ m paracord can be used as the semi-infinite string. Longer semi-infinite segment could also be employed, but, in this case, it would be better to change the orientation of the rope from horizontal to vertical to avoid sagging. 

As mentioned before, frequency bandgaps will be shown up as no oscillations in the last segment. After setting up the experiment, the frequency of the electric vibrator is varied, and measurements on the oscillations of the last segment will be performed, by using a high-speed camera (in a more expensive experimental setup) or by using stroboscopic light or opto-switch sensors (a less expensive one). The use of a high-speed camera is straightforward, and analysis can be conducted by using the freeware Tracker(for an introduction to video analysis of experimental results using this freeware, we refer the reader to Ref. \cite{vitor}). For low-cost alternatives, a stroboscopic light can be employed only for visual observation since it does not measure the oscillations on the last segment. On the other side, by calibrating an opto-switch sensor \cite{LeCarrou2014}, the oscillations can be measured, and quantitative analysis can be performed. This allows creating a plot that relates the oscillation amplitude with frequency, showing a bandgap for specific frequency ranges.


\end{document}